# Electron Spin Precession for the Time Fractional Pauli Equation

Hosein Nasrolahpour[*]

Department of Physics, Faculty of Basic Sciences,
University of Mazandaran,
P. O. Box 47416-95447, Babolsar, IRAN

Hadaf Institute of Higher Education, P.O. Box 48175 – 1666, Sari, Mazandaran, IRAN

**Abstract**

In this work, we aim to extend the application of the fractional calculus in the realm of quantum mechanics. We present a time fractional Pauli equation containing Caputo fractional derivative. By use of the new equation we study the electron spin precession problem in a homogeneous constant magnetic field.

***Keywords***: Fractional quantum mechanics; Time fractional Pauli equation; Electron spin precession

**1- Introduction**

Fractional derivatives provide a relevant tool for understanding the dynamics of the complex systems. Recently fractional calculus has found interest and applications in the domain of quantum mechanics and field theory (see [5] and refs. in). Fractional quantum mechanics (FQM) is in fact generalization of standard quantum mechanics to discuss quantum phenomena in fractal and complex systems. Up to now, several different approaches on FQM have been investigated [6-19].

In FQM, space fractional Schrödinger equation has been obtained by Laskin [9]. He constructed the fractional quantum mechanics using the Lévy path integral and considered some aspects of the space fractional quantum system. Space fractional Schrödinger equation similar to the standard one satisfies the Markovian evolution law. However, one can introduce the time fractional Schrödinger equation to describe non- Markovian evolution in quantum mechanics [10].

In Refs. [10, 11] some properties of time fractional Schrödinger equation have been studied. It is showed that Hamiltonian in this case is non-Hermitian and non-local in time. Furthermore, it is found [10] that for free particle case the wave function evolves in such a way that the value of total probability increases to the value $\frac{1}{\alpha^2}$ (where $\alpha$ is the order of derivative in the time fractional Schrödinger equation) in the limit $t \to \infty$. The probability structure of time fractional Schrödinger equation in the case of low-level fractionality [20, 21] also has been discussed in [11]. It has been proved that for the basis state the probability density increases initially, and then fluctuates around a constant value in later times.

---

[*] E-mail address:  hnasrolahpour@gmail.com

In the standard quantum mechanics, the well-known Pauli equation (the non-relativistic evolution equation of spin $\frac{1}{2}$ particles) is given by [1]

$$[\frac{1}{2m}(\hat{P}-\frac{e}{c}A)^2 + e\phi + \mu_B \hat{\sigma}.B]\Psi = i\hbar\frac{\partial\Psi}{\partial t} \qquad (1)$$

Where A is the three-component magnetic vector potential and $\phi$ is the electric scalar potential and $\Psi = \begin{pmatrix}\Psi_1\\\Psi_2\end{pmatrix}$ is the two component spinor wave function and $\mu_B = \frac{|e|\hbar}{2mc}$ is the so-called Bohr magneton.

As we mentioned above, quantum phenomena in disordered systems such as random fractal structure can be discussed by the fractional quantum mechanics. In disordered systems spatial and temporal fluctuations often obey Gaussian statistics due to the central limit theorem. However, it has been found in recent years that anomalous broad Lévy type distributions of the fluctuations can occur in many complex systems [24]. With this motivation we aim to present a fractional Pauli equation containing Caputo fractional derivative and study the electron spin precession within this framework.

In the following, fractional calculus [2-4] is briefly reviewed in Sec. 2. The time fractional Pauli equation is presented in Sec. 3. The time-dependent spin function and expectation values for the spin observables are obtained in terms of the Mittag-Leffler functions. The probability for spin-up and spin-down also are calculated in this section. In Sec. 4, we will present our conclusions.

**2- Fractional calculus**

*2.1. The Caputo fractional derivative operator*

Fractional calculus (FC) has a long history, when the derivative of order $\alpha = \frac{1}{2}$ has been described by Leibniz in a letter to L'Hospital in 1695 [3]. FC is the calculus of derivatives and integrals with arbitrary order, which unify and generalize the notions of integer order differentiation and n-fold integration, which have found many applications in recent studies to model a variety processes from classical to quantum physics [4-21]. In contrast with the definition of the integer order derivative, the definition of the fractional one is not unique. In fact, there exists several definitions including, Grünwald - Letnikov, Riemann - Liouville, Weyl, Riesz and Caputo for fractional order derivative. Fractional differential equations defined in terms of Caputo derivatives require standard boundary (initial) conditions. For this reason, in this paper we only use the Caputo fractional derivative. The left (forward) Caputo fractional derivative of a function f(t) is defined by

$$^c_0D^\alpha_t f(t) = \frac{1}{\Gamma(n-\alpha)}\int_0^t (t-\tau)^{n-\alpha-1} f^{(n)}(\tau)d\tau, \qquad \alpha>0, t>0 \qquad (2)$$

Where n is an integer number and $\alpha$ is the order of the derivative such that n-1< $\alpha$ <n and $f^{(n)}(\tau)$ denotes the n-th derivative of the function $f(\tau)$. Laplace transform to Caputo's fractional derivative gives

$$L\{^c_0D^\alpha_t f(t)\} = s^\alpha F(s) - \sum_{m=0}^{n-1} s^{\alpha-m-1} f^{(m)}(0) \qquad (3)$$

Where $F(s)$ is the Laplace transform of $f(t)$.

*2.2. The Mittag-Leffler function*

The Mittag-Leffler function is a generalization of the exponential function, which plays an important role in the fractional calculus. The Mittag-Leffler function is such a one-parameter function defined by the series expansion as

$$E_\alpha(z) = \sum_{k=0}^{\infty} \frac{z^k}{\Gamma(1+\alpha k)} \quad \alpha \in \mathbb{C}, \alpha > 0, z \in \mathbb{C} \tag{4}$$

And its general two-parameter representations is defined as

$$E_{\alpha,\beta}(z) = \sum_{k=0}^{\infty} \frac{z^k}{\Gamma(\beta+\alpha k)} \quad \alpha, \beta \in \mathbb{C}, \alpha, \beta > 0, z \in \mathbb{C} \tag{5}$$

Where $\mathbb{C}$ is the set of complex numbers and $\Gamma(\alpha)$ denotes the gamma function. Laplace transform for Mittag-Leffler function is very useful in solving fractional differential equations. The Laplace transformations for several Mittag-Leffler functions are summarized below

$$L\{E_\alpha(-\lambda t^\alpha)\} = \frac{s^{\alpha-1}}{s^\alpha + \lambda} \tag{6a}$$

$$L\{t^{\alpha-1}E_{\alpha,\alpha}(-\lambda t^\alpha)\} = \frac{1}{s^\alpha + \lambda} \tag{6b}$$

$$L\{t^{\beta-1}E_{\alpha,\beta}(-\lambda t^\alpha)\} = \frac{s^{\alpha-\beta}}{s^\alpha + \lambda} \tag{6c}$$

Where $s > |\lambda|^{\frac{1}{\alpha}}$.

## 3- Time fractional Pauli equation

As we mentioned in section 1, one can introduce the time fractional Schrödinger equation to describe non-Markovian evolution [*] in quantum mechanics. In this section we generalize the time fractional Schrödinger equation [12, 13] and obtain the following time fractional Pauli equation

$$[\frac{1}{2m}(\hat{P} - \frac{e}{c}A)^2 + e\phi + \mu_B \hat{\sigma}.B]\Psi = i\hbar_\alpha \frac{\partial^\alpha \Psi}{\partial t^\alpha} \quad 0 < \alpha \leq 1 \tag{7}$$

Where $\hbar_\alpha = M_P c^2 T_P^\alpha$ is a scaled Planck constant. The parameters $M_P$ and $T_P$ are Planck mass and Planck time, respectively, which are defined as

$$T_P = \sqrt{\frac{G\hbar}{c^5}} \quad , \quad M_P = \sqrt{\frac{c\hbar}{G}} \tag{8}$$

---

[*] Non-Markovian systems appear in many branches of physics, such as quantum optics, solid state physics, quantum information processing, and quantum chemistry (see for example the Ref. [22]).

Where G and c are the gravitational constant and the speed of light in vacuum, respectively.

We now use the Eq. (7) to discuss the electron spin precession problem in a homogeneous constant magnetic field. For this purpose, we first assume that, the electron is fixed at a certain location and its spin is the only degree of freedom. Also, let the magnetic field consist of a constant field $\vec{B}$ in the Z direction

$$\vec{B} = B_0 \hat{k}$$

Therefore, that part of the time fractional Pauli equation Eq. (7), which contains the spin yields

$$(i\hbar_\alpha)^c_0 D^\alpha_t \chi = \hbar_\alpha \omega^\alpha_L \hat{\sigma}_z \chi \qquad (9)$$

Where $\omega_L = -\dfrac{eB}{2mc}$ are the so-called Larmor frequency and the Pauli matrices for a spin $\dfrac{1}{2}$ particles are as below

$$\hat{\sigma}_x = \begin{pmatrix} 0 & 1 \\ 1 & 0 \end{pmatrix}, \quad \hat{\sigma}_y = \begin{pmatrix} 0 & -i \\ i & 0 \end{pmatrix}, \quad \hat{\sigma}_z = \begin{pmatrix} 1 & 0 \\ 0 & -1 \end{pmatrix}$$

Science the Hamiltonian of our system is a 2x2 matrix, the spin function in arbitrary time t must be written as a column matrix of two components and can be derived as below,

$$\chi_\alpha(t) = \begin{pmatrix} a(t) \\ b(t) \end{pmatrix} = \begin{pmatrix} e^{i\gamma} \cos\left(\frac{\theta}{2}\right) E_\alpha(-i(\omega_L t)^\alpha) \\ e^{i\delta} \sin\left(\frac{\theta}{2}\right) E_\alpha(i(\omega_L t)^\alpha) \end{pmatrix} \qquad (10)$$

Where $\gamma$ and $\delta$ are arbitrary phase constants.

Now, we able to calculate the expectation values for the observables $\hat{s}_x, \hat{s}_y, \hat{s}_z$. Then we will have

$$<\hat{s}_x>_{\alpha,t} = \frac{\hbar}{2}(\chi^\dagger \hat{\sigma}_x \chi) = \frac{\hbar}{4}\sin(\theta)[e^{i(\delta-\gamma)}(E_\alpha(i(\omega_L t)^\alpha))^2 + e^{-i(\delta-\gamma)}(E_\alpha(-i(\omega_L t)^\alpha))^2] \qquad (11)$$

$$<\hat{s}_y>_{\alpha,t} = \frac{\hbar}{2}(\chi^\dagger \hat{\sigma}_y \chi) = \frac{i\hbar}{4}\sin(\theta)[e^{-i(\delta-\gamma)}(E_\alpha(-i(\omega_L t)^\alpha))^2 - e^{i(\delta-\gamma)}(E_\alpha(i(\omega_L t)^\alpha))^2] \qquad (12)$$

$$<\hat{s}_z>_{\alpha,t} = \frac{\hbar}{2}(\chi^\dagger \hat{\sigma}_z \chi) = \frac{\hbar}{2}\cos(\theta)[E_\alpha(-i(\omega_L t)^\alpha) E_\alpha(i(\omega_L t)^\alpha)] \qquad (13)$$

Here we can see explicitly that as $\alpha \to 1$, Eq. (13) gives

$$<\hat{s}_z>_{\alpha=1,t} = \frac{\hbar}{2}\cos(\theta)[E_1(-i(\omega_L t)) E_1(i(\omega_L t))] = \frac{\hbar}{2}\cos(\theta) \qquad (14)$$

So we have $<s_z>_{\alpha=1,t} = <s_z>_{\alpha=1,t=0}$, as expected from the standard quantum mechanics. Furthermore, for the special case of $\alpha = \dfrac{1}{2}$, we have

$$<\hat{s}_z>_{\alpha=\frac{1}{2},t} = \frac{\hbar}{2}\cos(\theta)e^{-2\omega_L t}erfc(i\sqrt{\omega_L t})[2 - erfc(i\sqrt{\omega_L t})] \qquad (15)$$

Where $erfc(z)$ denoted the complementary error function [23], which is defined by

$$erfc(z) = \frac{2}{\sqrt{\pi}}\int_z^\infty e^{-t^2}dt \qquad (16)$$

Also by use of Eq. (10) we can calculate the probability for spin-up, $P_{\alpha\uparrow}$, and spin-down, $P_{\alpha\downarrow}$, at $t > 0$.

So we have

$$P_{\alpha\uparrow} = |a(t)|^2 = \cos^2(\frac{\theta}{2})[E_\alpha(-i(\omega_L t)^\alpha)E_\alpha(i(\omega_L t)^\alpha)] \qquad (17)$$

$$P_{\alpha\downarrow} = |b(t)|^2 = \sin^2(\frac{\theta}{2})[E_\alpha(-i(\omega_L t)^\alpha)E_\alpha(i(\omega_L t)^\alpha)]. \qquad (18)$$

## 4- Discussion and Conclusion

Fractional quantum mechanics is a generalization of traditional quantum mechanics to discuss quantum phenomena in disordered systems such as random fractal structure. In disordered systems spatial and temporal fluctuations often obey Gaussian statistics due to the central limit theorem. However, it has been found in recent years that anomalous broad Lévy type distributions of the fluctuations can occur in many complex systems [24]. With this motivation a time fractional Pauli equation is introduced.

In this paper we have studied the spin precession in the framework of fractional dynamics. For this purpose: first, we have presented the time fractional Pauli equation. Utilizing this equation we have derived the time-dependent spin function for our system. Then using the Eq. (10), we have obtained the expectation values for the observables $\hat{s}_x, \hat{s}_y, \hat{s}_z$. The following result can be easily deduced:

By use of the standard Pauli equation we have,

$$<\hat{s}> = \frac{\hbar}{2}(\cos(2\omega_L t + \delta - \gamma)\sin(\theta), \sin(2\omega_L t + \delta - \gamma)\sin(\theta), \cos(\theta)) \qquad (19)$$

Obviously, the above relation can be deduced from Eqs. (11-13) as $\alpha \to 1$ and we will have a conserved spin component in the field direction while the spin precesses around the z axis with twice the Larmor frequency $2\omega_L$ (we know that this is due to the gyromagnetic factor 2 of the spin). But as we can see from the Eq. (13) we have:

$$<\hat{s}_z>_{\alpha,t} \neq <\hat{s}_z>_{\alpha,t=0}, \quad 0 < \alpha < 1 \qquad (20)$$

These results show that spin component in the field direction $s_z$ is not conserved for the arbitrary case of $\alpha$. Therefore, spin precession phenomena is complicated in the framework of fractional dynamics(also as a result, we can consider modifications to the gyromagnetic factor of the spin in this framework). Furthermore, we have calculated the time dependent probability for spin-up, $P_{\alpha\uparrow}$, and spin-down, $P_{\alpha\downarrow}$ at $t > 0$, for the arbitrary case of $\alpha$ ($0 < \alpha < 1$). As a consequence of these results we can predict a transition from one spin state to the other spin state in our system, when time flows.

In this paper we only consider the case of $0 < \alpha < 1$, it is also interesting to consider the case of $1 < \alpha < 2$ for the Eq. (7) and derive new results. We hope to do this in future work.

Recently fractional calculus has found interest and applications in the context of relativistic quantum mechanics and field theory [25-32]. We hope to study some applications of fractional calculus in these interesting areas of physics, as well.

**Acknowledgement**

The author is grateful to Richard Herrmann for many useful discussions and explanations.